\documentclass[aps,
prd,
nofootinbib,
showpacs,
twocolumn,
superscriptaddress,
preprintnumbers,
floatfix]{revtex4-2}

\usepackage[bookmarks, linktocpage, colorlinks = true, linkcolor = blue, urlcolor  = blue, citecolor = blue, anchorcolor = green, hyperindex = true, hyperfigures]{hyperref}

\usepackage{graphicx} 
\usepackage{dcolumn}
\usepackage{multirow}
\usepackage{amsmath, amssymb}
\usepackage{slashed}
\usepackage[usenames]{color}
\usepackage{float}
\usepackage{bm}

\newcommand{\Tc}{T_{\mathrm{c}}}
\newcommand{\Td}{T_{\mathrm{d}}}
\newcommand{\Th}{T_{\mathrm{H}}}

\begin{document}

\preprint{RIKEN-iTHEMS-Report-25}

\title{Description of the baryon mass spectrum by open strings and diquarks}

\author{Yuki Fujimoto}
\email{fujimotophy@gmail.com}
\affiliation{Department of Physics, Niigata University, Ikarashi, Niigata 950-2181, Japan}
\affiliation{Center for Interdisciplinary Theoretical and Mathematical Sciences (iTHEMS), RIKEN, Wako 351-0198, Japan}

\date{\today}

\begin{abstract}
    We analyze the mass spectra of hadrons and demonstrate that the physical spectra of mesons and baryons are well described by the exponential spectrum of open strings.
    The open-string spectrum, derived from string theory, is characterized by a unique Hagedorn temperature $T_{\rm H}$ and free from any other parameters.
    Notably, our fitting to the physical spectra yields consistent values for both mesons and baryons, $T_{\rm H} \simeq 0.34\,\text{GeV}$, which contrasts with previous phenomenological analyses that suggested different values.
    This obtained value aligns well with typical string tension derived from lattice-QCD calculations and the Regge slope.
    In the baryonic sector, our results indicate that diquarks play a crucial role in describing the mass spectrum, implying that baryons can be understood as a quark-diquark system, as anticipated by Regge phenomenology.
    These findings have significant implications for our understanding of quark deconfinement, especially in the possibly existing regime at high temperature and small baryon chemical potential within the QCD phase diagram.
\end{abstract}

\maketitle

\section{Introduction}

It is widely believed that deconfinement from hadronic matter to quark-gluon plasma (QGP) occurs at a temperature $T$ close to the chiral crossover temperature, around $\Tc \simeq 0.16\,\text{GeV}$~\cite{Aoki:2006we,HotQCD:2018pds,Borsanyi:2020fev} (see, e.g., Ref.~\cite{Fukushima:2017csk} for review).
However, recent lattice-QCD studies suggest a different view on the deconfinement.
For example, the meson state persists even above $\Tc$ for both heavy and light flavors~\cite{Rothkopf:2019ipj,Petreczky:2021zmz,Lowdon:2022xcl,Aarts:2023nax,Bala:2023iqu}, and in connection with that, the real part of the heavy quark-antiquark potential at finite $T$ remains unscreened at least up to $2\Tc$~\cite{Bazavov:2023dci, Bala:2025ilf} (see also Refs.~\cite{Asakawa:2003re,Rothkopf:2011db,Burnier:2013nla,Burnier:2015tda,Bala:2019cqu,Bala:2021fkm,Spriggs:2023ccb}).
Further, there is evidence for the second transition around $2\Tc$ up to where the center-vortex percolation persists~\cite{Mickley:2024vkm}.

These observations have led to the speculation that there may be a novel region in the QCD phase diagram between hadronic matter and QGP~\cite{Rohrhofer:2019qwq, Rohrhofer:2019qal, Glozman:2022lda, Glozman:2022zpy, Cohen:2023hbq, Fujimoto:2025sxx, Hanada:2025rca}.
This region, dubbed spaghetti of quarks with glueballs (SQGB), shows distinct deconfinement properties from QGP and incorporates the color flux tubes and glueballs as relevant degrees of freedom.
The multiplicity of glueballs is scarce in the SQGB, so flux tubes dominate and it becomes stringy fluid.
In the SQGB, quarks should be understood as being connected by these open color strings.
These quarks appear to be thermally liberated because it is energetically more favorable to form a new open string with a nearby meson than to stretch the original string.
This aligns with the behavior of (correlators of) the Polyakov loop~\cite{Bazavov:2020teh}.

That the SQGB embodies open strings is supported by the fact that QCD thermodynamics are well reproduced by ideal gas of open strings~\cite{Fujimoto:2025sxx}.
In the usual treatment, the thermodynamic properties of the low-$T$ QCD matter can largely be accounted by non-interacting hadron resonance gas model.
Thus, this in turn suggests that the hadron mass spectrum can well be described by the open-string spectrum, and it was confirmed in Ref.~\cite{Marczenko:2025nhj} that this is indeed the case.
The open-string density of states can be derived from string theory and predicts an exponentially growing mass spectrum; the exponent is characterized by the Hagedorn limiting temperature $\Th$.
The string spectrum looks similar to the Hagedorn spectrum introduced in the context of statistical bootstrap model~\cite{Hagedorn:1965st,Frautschi:1971ij}, but theory is fundamentally different from the bootstrap model.
In string theory, $\Th$ is related to the string tension $\sigma$ or the Regge slope $\alpha' = 1/ (2\pi \sigma)$ and, therefore, to deconfinement.
In fact $\Th$ can be interpreted as the temperature at which partons deconfine~\cite{Cabibbo:1975ig} (see also Ref.~\cite{Fujimoto:2021xix} for application).

In Ref.~\cite{Fujimoto:2025sxx}, it was found that the value of $\Th$ that best reproduces the QCD thermodynamics is $\Th \simeq 0.3\,\text{GeV}$.
This is close to the deconfinement temperature $\Td \simeq 0.285 \, \text{GeV}$ in pure SU(3) Yang-Mills theory~\cite{Borsanyi:2022xml, Giusti:2025fxu, Svetitsky:1985ye} and almost twice the chiral pseudocritical temperature $\Tc$.
Incidentally, this description using the string spectrum was inspired by the closed-string description of glueball gases in pure Yang-Mills theory, in which the author also found $\Th \simeq 0.3 \,\text{GeV}$~\cite{Meyer:2009tq}.
The use of string-based spectrum in the mesonic sector led to the increase in $\Th$.
The typical value obtained by using the Hagedorn spectrum from the statistical bootstrap is around $\Th \simeq 0.15\,\text{GeV}$ (See, e.g., Refs.~\cite{Broniowski:2000bj,Broniowski:2004yh,Majumder:2010ik,Lo:2015cca,ManLo:2016pgd,Rafelski:2015xej}).
The notable difference is that while the spectrum in the statistical bootstrap model suffers from the ambiguity in the coefficient, the string theory determines the coefficient without ambiguity, leading to more robust prediction.
We note, however, that such string-based description used in Refs.~\cite{Fujimoto:2025sxx,Marczenko:2025nhj} was limited to mesons.

In this work, we extend the string-based description to baryon spectrum.
In doing so, we regard a baryon as a system of a quark-diquark pair.
A diquark is strong correlation between quarks inside hadron.
This is not a gauge-invariant quantity so is not a well-defined object appearing in the asymptotic state, nevertheless, it is supported by many successful phenomenologies by assuming a diquark as a constituent of hadrons (see, e.g., Refs.~\cite{Jaffe:2004ph,Shifman:2024kfj} for review).
Among such successful applications of diquarks, the baryonic states lie on the straight-line Regge trajectories if one assumes diquark correlation inside baryons~\cite{Wilczek:2004im,Selem:2006nd}.
This can be interpreted as arising from a rotating string connecting a quark-diquark configuration.
Also, this picture is in accordance with the classical string models that the Y-shaped baryon junction is unstable against the collapse to the asymmetric quark-diquark junction~\cite{Sharov:2000dt,Sharov:2000pg,tHooft:2004doe}.
One may also consider a possibility of the closed $\Delta$-shaped baryon with quarks on each vertex, but this does not work because the closed-string density of states is much suppressed compared to the open string~\cite{Fujimoto:2025sxx}.

\section{Mass spectra of hadrons and open strings}

For hadron mass spectra, we use the 2020 edition of the Particle Data Group (PDG) tables~\cite{ParticleDataGroup:2020ssz}, which we adopted from the latest version of Thermal-FIST~\cite{Vovchenko:2019pjl}.
The table only includes established mesons and baryons with three- and four-star ratings in the PDG.
All the states are approximately treated as stable point-like particles.
The density of states for the discrete hadron spectrum is given by
\begin{align}
    \rho(m) = \sum_i d_i \delta(m- m_i)\,,
\end{align}
where the sum runs over all stable hadrons and resonances with mass $m_i$ and degeneracy factor $d_i$.
It is customary to consider the cumulative mass spectrum~\cite{Broniowski:2000bj}:
\begin{align}
    N(m) = \int_0^m dm' \,\rho(m')\,,
\end{align}
For the density of states for the discrete spectrum, the cumulative spectrum is given by a step function
\begin{align}
    N(m) = \sum_i d_i \theta(m - m_i)\,.
\end{align}
As shown in Figs.~\ref{fig:meson} and \ref{fig:baryon}, the spectrum shows exponential growth.
Such growth is predicted in Hagedorn's statistical bootstrap model and usually fitted with the following equation~\cite{Hagedorn:1965st}:
\begin{align}
    \rho_{\mathrm{H}}(m) = \frac{c}{(m^2 + m_0^2)^{a/2}}e^{m/\Th}\,.
    \label{eq:rhoH}
\end{align}
This equation has three free parameters $a$, $c$, and $m_0$.
Typically, for $2.5 \leq a \leq 4$, the fitted value of $\Th$ ranges between $[0.14, 0.16]\,\text{GeV}$ (see Chap.~22 in Ref.~\cite{Rafelski:2015xej}).

String theory is also known to naturally predict exponentially increasing mass spectrum from the partition of an integer number appearing in quantization of strings.
The asymptotic open bosonic string density of states spectrum in $D$ spacetime dimensions is given by~\cite{Green:1987sp,Fujimoto:2025sxx}
\begin{align}
    \rho_{\mathrm{str}} (m) 
    &= \frac{(b\pi/6)^{-b/2}}{\sqrt{2\pi} \Th}  \left(\frac{m}{\Th}\right)^{-(b+1)/2} e^{m/\Th}\,, \label{eq:rhostrb}\\
    &\overset{D=4}{=} \frac{\sqrt{2\pi}}{6 \Th} \left(\frac{m}{\Th}\right)^{-3/2} e^{m/\Th}\,,
    \label{eq:rhostr}
\end{align}
where $b=D-2$ is the number of polarizations.
This is the open-string counterpart of the closed-string formula used in Ref.~\cite{Meyer:2009tq}.
The string formula is free from the ambiguity in the coefficient as in Eq.~\eqref{eq:rhoH}.
The only free parameter is the Hagedorn temperature, which is related to the tension of the confining string $\sigma$ through the following relation:
\begin{align}
    \Th = \sqrt{\frac{3\sigma}{b\pi}} \overset{D=4}{=} \sqrt{\frac{3\sigma}{2\pi}}\,.
    \label{eq:TH}
\end{align}
Notably, if one uses the oft-used value $\sigma \simeq 1 \,\text{GeV/fm}$ in this formula, one gets $\Th \simeq 0.31 \, \text{GeV}$.
Also, the typical Regge slope $\alpha' \simeq 1 \,\text{GeV}^{-2}$ leads to a similar value $\Th \simeq 0.28 \, \text{GeV}$.

Note the difference between the previous result obtained in Ref.~\cite{Huang:1970iq} within dual resonance model~\cite{Veneziano:1974dr} and ours.
In Ref.~\cite{Huang:1970iq}, the authors obtained a $-5/2$ power of $m$ in the coefficient, whereas it is $-3/2$ in our result;
this exponent $-5/2$ by coincidence matched with the exponent in the Hagedorn's original work~\cite{Hagedorn:1965st}, corresponding to taking $a=5/2$ in Eq.~\eqref{eq:rhoH}.
This is because they substituted $D$ instead of $b$ in Eqs.~\eqref{eq:rhostrb} and \eqref{eq:TH}.

\section{String description of meson spectrum}

Here, we review the open-string description of the meson mass spectrum.
We describe a meson as an open color flux tube with a constituent quark and a constituent antiquark attached at its each end.
Such description is already known to work well~\cite{Fujimoto:2025sxx,Marczenko:2025nhj}.

The cumulative mass spectrum of mesons is
\begin{align}
    N(m) &= \sum_{i = \pi,K,\eta} d_i \theta(m - m_i) \notag \\
    &\qquad + \int_0^m dm' \, d_M(m')\rho_{\mathrm{str}}(m') \,.
\end{align}
We separated out the pseudo-Nambu-Goldstone (NG) mesons, which are $\pi$, $K$, and $\eta$, as they are exceptionally light compared to other mesons.

The string degeneracy factor $d_M$ for mesons, appearing in the open-string spectrum, varies with the mass range:
\begin{align}
    d_M(m) = \sum_i d_{M,i} \theta (m - m_{M,i})\,.
\end{align}
The values of $d_{M,i}$ and $m_{M,i}$ are summarized in Table~\ref{tab:meson}.
\begin{table}[H]
    \centering
    \begin{tabular}{ccc}
    \hline
    Composition $i$ & Degeneracy $d_{M,i}$ & Threshold $m_{M,i}$ [GeV] \\
    \hline
    $l\mbox{--}\bar{l}$ & $(2N_l)^2 = 16 $ & $m_{\rho(770)} \simeq 0.775$ \\
    $l\mbox{--}\bar{s}, s\mbox{--}\bar{l}$ & $4(2N_l)= 16 $ & $m_{K^\ast(892)} \simeq 0.896$ \\
    $s\mbox{--}\bar{s}$ & $4$ & $m_{\phi(1020)} \simeq 1.019$ \\
    \hline
    \end{tabular}
    \caption{Parameters in the open-string degeneracy factor for mesons. See text for explanation.}
    \label{tab:meson}
\end{table}

In the table, $l$ refers to light $u$ and $d$ quarks, and $N_l = 2$ is the number of light flavors.
The degeneracy factor $d_{M,i}$ arises from spin and flavor of (anti)quarks attached to the end of an open string.
The threshold mass $m_{M,i}$ is taken to be the mass of the lightest meson apart from the pseudo-NG bosons.
For the composition $l$--$\bar{l}$, the lightest non-NG meson is $\rho(770)$ whose mass is $m_{\rho(770)} \simeq 0.775\,\text{GeV}$.
Likewise, for the compositions $l$--$\bar{s}$ and $s$--$\bar{l}$ with strangeness $|S|=1$, the lightest non-NG meson is $K^\ast(892)$, and for $s$-$\bar{s}$ with $|S|=2$, the lightest is $\phi(1020)$.
In the preceding work, these values were taken to be the sum of the constituent quark mass, but here we take it as a meson mass to avoid ambiguity associated with the extraction of the constituent quark mass.

\begin{figure}[t]
    \centering
    \includegraphics[width=0.95\columnwidth]{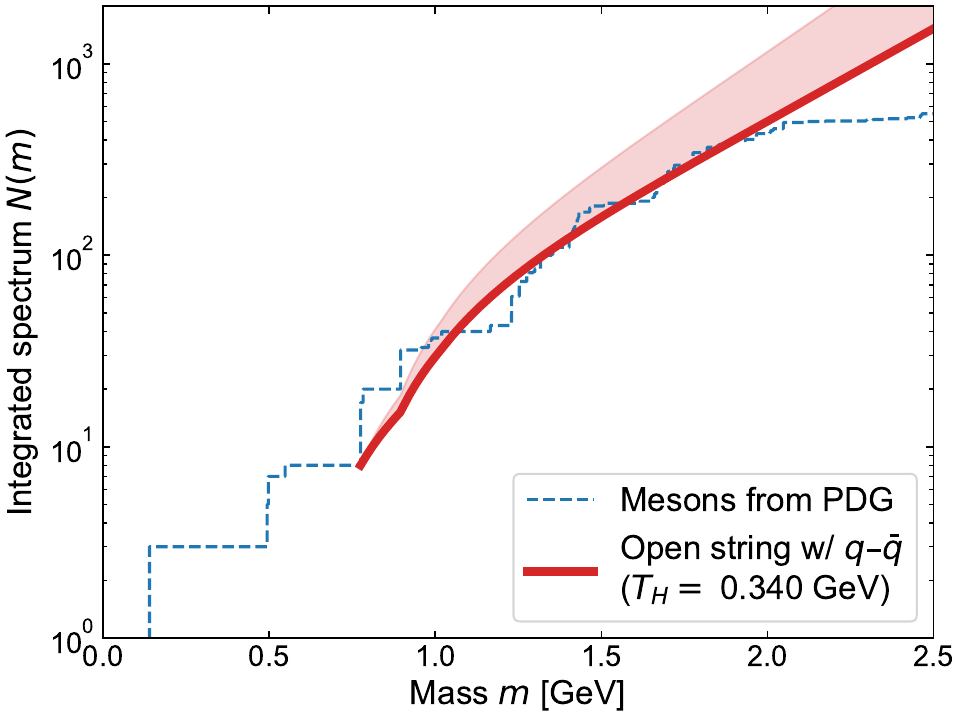}
    \caption{The meson mass spectrum from PDG (blue dashed line) and the curve fitted by the open string formula (red curve).
    The shaded band corresponds to the varying values of $\Th$ in the range $[0.285, 0.340]\,\text{GeV}$.}
    \label{fig:meson}
\end{figure}

In Fig.~\ref{fig:meson}, we compare the PDG meson spectrum and the open-string spectrum.
As we constructed above, the open-string spectrum has a single free parameter, $\Th$.
One can therefore perform non-linear least-squares fit of this value to the PDG meson spectrum.
By fitting in the mass range $m_{\rho(770)} < m < 1.5\,\text{GeV}$, we obtain a value $\Th \simeq 0.340\,\text{GeV}$.
This value is more than twice as large as the conventional $\Th$ extracted by using Eq.~\eqref{eq:rhoH} and rather is close to the deconfinement temperature in the pure SU(3) Yang-Mills theory, $\Td \simeq 0.285 \,\text{GeV}$~\cite{Borsanyi:2022xml,BMW:2012hcm}.

\section{String description of baryon spectrum}

Here, we explore the possibility of describing the baryonic mass spectrum below 2\,GeV by an open string, given the successful description of mesonic mass spectrum by an open string.
As we already noted in the introduction, this is motivated by the fact that the baryonic spectrum are found to lie on a linear Regge trajectory with a quark-diquark configuration~\cite{Wilczek:2004im,Selem:2006nd}.

Since each quark is a color triplet, a diquark can form an antisymmetric color $\bm{\bar{3}}_c$, or a symmetric color $\bm{6}_c$ representation.
Normally, a color $\bm{\bar{3}}_c$ diquark is solely considered because this channel is attractive by one-gluon exchange, and the color $\bm{6}_c$ is unfavored configuration due to large electrostatic field energy.
Thus, in terms of the color representation, the QCD string does not distinguish between an antiquark and the color $\bm{\bar{3}}_c$ diquark.
This justifies the open-string description of baryons as a quark-diquark system.

The Fermi statistics require the total diquark wave function to be antisymmetric.
Since color $\bm{\bar{3}}_c$ is antisymmetric representation, spin and flavor must be symmetrized in a diquark.
One can either antisymmetrize or symmetrize both indices.
It turns out that spin antisymmetric configuration is favored over the symmetric one.
This is due to the spin-dependent chromomagnetic interaction, derived from the nonrelativistic expansion of the one-gluon exchange amplitude.
This spin antisymmetric diquark is spin singlet and flavor antisymmetric (for the non-strange case, this is isosinglet $I=0$) configuration.
We denote such favored ``good'' diquark configuration as $[qq]$.
One can also consider spin triplet and flavor symmetric (isotriplet $I=1$ for light quarks), and we write this ``bad'' diquark configuration as $(qq)$.

As in the case of mesons, we consider the cumulative mass spectrum for baryons:
\begin{align}
    N(m) = \int_0^m dm' \, d_B(m')\rho_{\mathrm{str}}(m') \,,
\end{align}
where the string mass spectrum is the same as in Eq.~\eqref{eq:rhostr}.
The string polarization degeneracy factor $d_B(m)$ for baryons is given by
\begin{align}
    d_B(m) = \sum_i d_{B,i} \theta (m - m_{B,i})\,.
\end{align}
The values of $d_{B,i}$ and $m_{B,i}$ are summarized in Table~\ref{tab:baryon}.
\begin{table}[H]
    \centering
    \begin{tabular}{cccc}
    \hline
    Composition $i$ & $(I,S)$ & $d_{B,i}$ & $m_{B,i}$ [GeV] \\
    \hline
    $l\mbox{--}[ll]$ & $(\frac12, 0)$ & $4$ & $m_{p} \simeq 0.938$ \\
    $s\mbox{--}[ll]$ & $(0, -1)$ & $2$ & $m_{\Lambda} \simeq 1.116$ \\
    $l\mbox{--}[ls]$ & $(1, -1)$ & $6$ & $m_{\Sigma^+} \simeq 1.189$ \\
    $l\mbox{--}(ll)$ & $(\frac32, 0)$ & $16$ & $m_{\Delta} \simeq 1.232$ \\
    $s\mbox{--}[ls]$ & $(\frac12, -2)$ & $4$ & $m_{\Xi^0} \simeq 1.315$ \\
    $s\mbox{--}(ss)$ & $(0, -3)$ & $4$ & $m_{\Omega^-} \simeq 1.672$ \\
    \hline
    \end{tabular}
    \caption{Parameters in the open-string degeneracy factor for baryons. See text for explanation.}
    \label{tab:baryon}
\end{table}

One can consider multiple combinations for the composition of a baryon as a quark-diquark system.
For a single quark at one end, one can have an $l$ or $s$ quark.
For a diquark at the other end, one can consider the multiple possibilities: a good diquark comprising $[ll]$ or $[ls]$, or a bad diquark comprising $(ll)$, $(ls)$, or $(ss)$.
Note that the $ss$ diquark can only form the bad diquark configuration as it is inevitably symmetric in flavor.

Let us construct a possible baryonic states out of these constituents.
States we consider are listed in Table~\ref{tab:baryon}, and each state is labeled by the isospin and strangeness $(I,S)$.
The open string does not modify these quantum numbers, so one has to include quark-diquark pairs with all the possible combinations of $(I,S)$ to reproduce all the baryonic states.
In contrast, a quark-diquark pair does not have to account for all the spin states;
the total angular momentum $J$ of baryon in our description can take any value since the orbital motion of the open string can induce a tower of different angular momentum states.
Therefore, in the following, we do not include configurations involving the bad diquarks such as $l$--$(ls)$ and $s$--$(ls)$, which correspond to the decuplet states $\Sigma^*(1385)$ and $\Xi^+(1530)$, respectively.
The only bad diquark configurations that we include are $l$--$(ll)$ and $s$--$(ss)$, since these correspond to $(I,S)$ quantum numbers that cannot be reproduced by the good diquark configurations.

Now, we explain the counting of $d_{B,i}$ in Table~\ref{tab:baryon}.
First, we consider the composition $l$--$[ll]$, corresponding to nucleons.
The degeneracy of this composition is $4$ due to a total spin and isospin of $1/2$.
Next, for the composition $s$--$[ll]$, corresponding to $\Lambda$'s, the degeneracy is $2$ because the total spin is $1/2$ and the total isospin is $0$.
Then, we examine the $l$--$[ls]$ composition, which corresponds to $\Sigma$'s.
Here, the total spin remains $1/2$, but the total isospin can be $0 \oplus 1$ via $\frac12 \otimes \frac12$.
The $I=0$ and $I=1$ states corresponds to $\Lambda$'s and $\Sigma$'s, respectively.
Since the $\Lambda$ state is already included in the $s$--$[ll]$ configuration, we only consider the $I=1$ state to avoid double counting.
We assign a factor $6$ for the spin-isospin degeneracy of the $l$--$[ls]$ configuration.
As explained below, this exclusion has a minor effect on the fitted Hagedorn temperature.
Next, we look at the $l$--$(ll)$ composition with a bad diquark, corresponding to $\Delta$'s.
The degeneracy requires careful consideration since both total spin and isospin can be $3/2$ or $1/2$, following $1 \otimes \frac12 = \frac32 \oplus \frac12$.
For baryons, the antisymmetry of the total wave function necessitates a symmetric spin-isospin combination, which is possible only for $(3/2, 3/2)$ (corresponds to $\Delta$'s) and $(1/2, 1/2)$ (for nucleons).
As nucleons are already covered in the $l$--$[ll]$ configuration, we retain only the $I=3/2$ state to avoid double counting and assign a factor of 16 for the spin-isospin degeneracy of this configuration.
Then, we analyze the $s$--$[ls]$ composition, corresponding to $\Xi$'s, which has a degeneracy of $4$ due to a total spin and isospin of $1/2$.
Finally, for the $s$--$(ss)$ composition, corresponding to $\Omega$'s, the degeneracy is 4 due to a total spin of $3/2$.
For each configuration, we use the mass of the lightest baryon with the same $(I,S)$ value as the mass threshold. 

\begin{figure}[t]
    \centering
    \includegraphics[width=0.95\columnwidth]{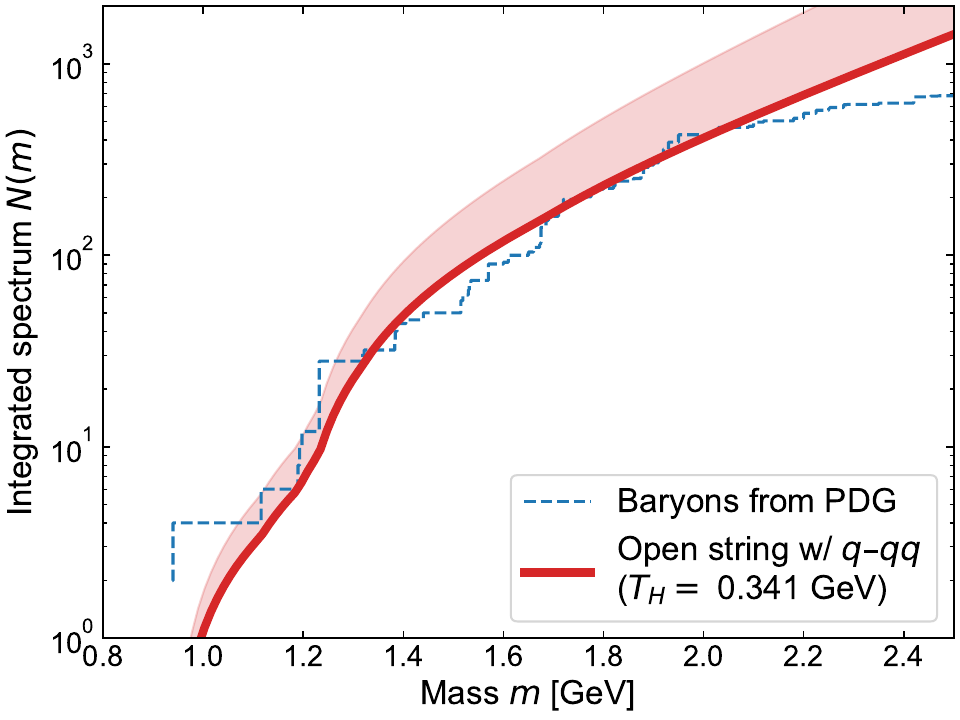}
    \caption{The baryon mass spectrum from PDG (blue dashed line) and the curve fitted by the open-string formula (red curve).
    The shaded band corresponds to the varying values of $\Th$ in the range $[0.285, 0.340]\,\text{GeV}$.}
    \label{fig:baryon}
\end{figure}

In Fig.~\ref{fig:baryon}, we compare the PDG baryon spectrum and the open-string spectrum with a quark-diquark configuration.
We fit $\Th$ in the open-string spectrum to the PDG baryon spectrum in the mass range $m_{p} < m < 2.0\,\text{GeV}$.
We obtain a value $\Th \simeq 0.341\,\text{GeV}$.
Surprisingly, this value is almost the same as $\Th$ extracted from the meson spectrum.
This strongly supports that there is only one deconfinement scale in QCD.
In other words, quarks deconfine at the same energy scale as gluons regardless of whether they are in mesons or baryons.
Also, this is consistent with the observation that the values of the Regge slope read out from the baryonic and the mesonic Regge trajectories do not differ from each other~\cite{Selem:2006nd}.

We note that even if one keeps the doubly counted states in the configurations $l$--$[ls]$ and $l$--$(ll)$, i.e.\ take their degeneracy factor as $8$ and $20$, respectively, the fitted Hagedorn temperature becomes $\Th \simeq 0.354\,\text{GeV}$, which only amounts to a few percent change in the fitted value.

The values of $\sigma$ determined in the recent literature are $\sqrt{\sigma} = 0.485(6)\,\text{GeV}$~\cite{Athenodorou:2020ani} and $\sqrt{\sigma} = 0.4817(97) \,\text{GeV}$~\cite{Brambilla:2022het}.
These corresponds to the Hagedorn temperature $\Th = 0.335(4)\,\text{GeV}$ and $\Th = 0.3329(67)\,\text{GeV}$.
These results are consistent with our current estimate of $\Th$.

\section{Summary and outlook}

We presented a semi-quantitative analysis of baryon mass spectrum using the mass spectrum derived from string theory.
Our description involves minimal modeling in the way that only the quark-diquark composition of baryons and their threshold are modeled to reproduce all the isospin and strangeness states in the physical spectrum.
Threshold mass is fixed to the observed values, and the only free parameter is the Hagedorn temperature $\Th$.
By fitting $\Th$ in the string-based spectra to the physical mass spectra, we found the common values $\Th \simeq 0.34\,\text{GeV}$ for both mesonic and baryonic cases.
The value is consistent with zero-temperature lattice-QCD calculation, and also with an observation that the baryons and mesons have the same Regge slope in the Chew-Frautschi plot.
Our result further adds supporting argument toward the diquark correlation inside baryons.

With this handy formula of baryonic states at hand, one can now extend our string-model-based study of the QCD phase diagram to the finite chemical potential region.
As a first step, applications to QCD thermodynamics can be considered.
One can also test this behavior by extending to the higher-order cumulants coupled to the chemical potentials.

\begin{acknowledgments}
    I would like to thank Larry McLerran for useful and encouraging correspondences.
    This work is supported by JSPS KAKENHI Grant-in-Aid for Research Activity Start-up (No.~25K23388).
\end{acknowledgments}

\bibliography{diquark}
\end{document}